\documentclass[12pt,a4paper]{article}
\usepackage{graphicx,hyperref,wrapfig,cite,color,subfigure,amssymb,amsmath}
\hypersetup{
     colorlinks=true,       
     linkcolor=red,          
     citecolor=green,        
     filecolor=magenta,      
     urlcolor=blue           
}
\newcommand{\newc}{\newcommand}
\def\u#1{\verb!#1!\endgroup}

\newc{\HW}{\textsf{HERWIG}}
\newc{\TAUOLA}{\textsf{TAUOLA}}
\newc{\ThePEG}{\textsf{ThePEG}}
\newc{\HWPP}{\textsf{Herwig++}}
\newc{\evt}{\textsf{EvtGen}}
\newc{\fortran}{\textsf{FORTRAN}}
\newc{\decayer}{\textsf{Decayer}}

\newc{\HWPPClass}[1]{\href{http://projects.hepforge.org/herwig/doxygen/classHerwig_1_1#1.html}{\textsf{#1}}}
\newc{\ThePEGClass}[1]{\href{http://projects.hepforge.org/thepeg/doxygen/classThePEG_1_1#1.html}{\textsf{#1}}}
\newc{\HWPPParameter}[2]{\href{http://projects.hepforge.org/herwig/doxygen/#1Interfaces.html\##2}{{\bf #2}}}
\newc{\ThePEGParameter}[2]{\href{http://projects.hepforge.org/thepeg/doxygen/#1Interfaces.html\##2}{{\bf #2}}}
\newc{\HWPPParameterValue}[3]{\href{http://projects.hepforge.org/herwig/doxygen/#1Interfaces.html\##2}{{\bf [#2=#3]}}}
\newc{\HWPPParameterValueB}[3]{\href{http://projects.hepforge.org/herwig/doxygen/#1Interfaces.html\##2}{{\bf [#3]}}}
\newc{\ThePEGParameterValue}[3]{\href{http://projects.hepforge.org/thepeg/doxygen/#1Interfaces.html\##2}{{\bf [#2=#3]}}}

\begin{document}
\tolerance=100000
\thispagestyle{empty}
\setcounter{page}{0}
 \begin{flushright}
IPPP/11/09\\
MCnet-11-04\\
DCPT/11/18\\
DESY 11-018\\
KA-TP-04-2011

\end{flushright}
\begin{center}
{\Large \bf Herwig++ 2.5 Release Note}\\[0.7cm]

S.~Gieseke$^1$,
D.~Grellscheid$^2$,
K.~Hamilton$^3$,
A.~Papaefstathiou$^4$,
S.~Pl\"atzer$^5$,
P.~Richardson$^{2}$,
C.~A.~R\"ohr$^1$,
P. Ruzicka$^6$,
A.~Si\'odmok$^1$,
L.~Suter$^{7}$,
D.~Winn$^2$

E-mail: {\tt herwig@projects.hepforge.org}\\[1cm]

$^1$\it Institut f\"ur Theoretische Physik, Karlsruhe Institute of Technology.\\[0.4mm]
$^2$\it IPPP, Department of Physics, Durham University. \\[0.4mm]
$^3$\it INFN, Sezione di Milano-Bicocca.\\[0.4mm] 
$^4$\it Cavendish Laboratory, University of Cambridge.\\[0.4mm]
$^5$\it Theory Group, DESY Hamburg.\\[0.4mm]
$^6$\it Institute of Physics, Academy of Sciences of the Czech Republic.\\[0.4mm]
$^7$\it School of Physics and Astronomy, University of Manchester.
\end{center}

\vspace*{\fill}

\begin{abstract}{\small\noindent A new release of the Monte Carlo
    program \HWPP\ (version 2.5) is now available. This version comes with
    a number of improvements including: 
    new next-to-leading order matrix elements, including weak boson pair production;
    a colour reconnection model; diffractive processes; additional
    models of physics beyond the Standard Model; new leading-order
    matrix elements for hadron--hadron and
    lepton--lepton collisions as well as photon-initiated processes.
  }
\end{abstract}

\tableofcontents
\setcounter{page}{1}

\section{Introduction}

The last major public version (2.3) of \HWPP, is described in great detail
in \cite{Bahr:2008pv,UpdatedManual,Bahr:2008tx,Bahr:2008tf}. As we did not produce a release note for version 2.4 we
describe all changes since version 2.3 in this release note.
The manual will be updated to 
reflect these changes and this release note is only intended to highlight these
new features and the other minor changes made since version~2.3.

Please refer to \cite{Bahr:2008pv} and the present paper if
using version 2.5 of the program.

The main new features of this version are: new next-to-leading order~(NLO) matrix elements, including weak boson pair production;
a colour reconnection model; diffractive processes; additional
models of physics beyond the Standard Model; new leading-order matrix
elements for hadron--hadron and
lepton--lepton collisions as well as photon-initiated processes. 

In addition, the \textsf{MC@NLO}~\cite{Frixione:2010ra} program
can now generate partonic configurations which can be showered and hadronized using \HWPP\ to
produce events with next-to-leading-order accuracy in the \textsf{MC@NLO} approach to matching NLO matrix elements
and the parton shower.

\subsection{Availability}

The new program version, together  with other useful files and information,
can be obtained from the following web site:
\begin{quote}\tt
       \href{http://projects.hepforge.org/herwig/}{http://projects.hepforge.org/herwig/}
\end{quote}
  In order to improve our response to user queries, all problems and requests for
  user support should be reported via the bug tracker on our wiki. Requests for an
  account to submit tickets and modify the wiki should be sent to 
  {\tt herwig@projects.hepforge.org}.

  \HWPP\ is released under the GNU General Public License (GPL) version 2 and 
  the MCnet guidelines for the distribution and usage of event generator software
  in an academic setting, which are distributed together with the source, and can also
  be obtained from
\begin{quote}\tt
 \href{http://www.montecarlonet.org/index.php?p=Publications/Guidelines}{http://www.montecarlonet.org/index.php?p=Publications/Guidelines}
\end{quote}

\section{POWHEG}

  The previous version of \HWPP\ included for the first time a number of
  processes simulated according to the POWHEG NLO 
  parton shower matching scheme~\cite{Nason:2004rx,Frixione:2007vw}.  
  The current release builds on this internal library of NLO-accurate
  POWHEG simulations, adding doubly resonant $W^{+}W^{-}$, $W^{\pm}Z^0$ and
  $Z^0Z^0$ pair production processes, fermionic Higgs decays and $e^+e^-\to q\bar q$.

  This release also sees a major restructuring of the code for the parton shower,
  in particular the code handling the real emission in the POWHEG scheme,
  in order to make the implementation of further processes in the POWHEG
  scheme and other extensions in the future easier.

Previously, the hard and soft matrix element
corrections were implemented in a set of dedicated \textsf{MECorrection} classes
inheriting from the \HWPPClass{MECorrectionBase} class. Similarly, the real emission 
corrections in the POWHEG scheme were implemented in a set of \textsf{HardGenerator}
classes inheriting from the \HWPPClass{HardestEmissionGenerator} base class. While
this had some advantages, in particular allowing the corrections to be applied regardless of
how the hard processes was generated (for example the corrections could still be
used with events read from a LHE file), it led to a significant 
replication of code between the \textsf{HardGenerator} and matrix element
classes implementing the $\bar{B}$ function in the POWHEG scheme. We have therefore 
restructured the code so that both the matrix element corrections and real corrections
in the POWHEG scheme are implemented in the matrix element or \textsf{Decayer} class
implementing the hard scattering or decay process.
In addition, the functionality of the separate \HWPPClass{PowhegEvolver} has been merged
into the base \HWPPClass{Evolver} class.

\subsection{Vector Boson Pair Production}

\begin{figure}
\begin{centering}

  \subfigure[Cosine of the polar angle
      of a positron produced from a decaying $Z^0$ boson, in its rest frame,
      in $Z^0Z^0$ pair production at the Tevatron ($\sqrt{s}=1.96$~TeV). 
      The \HWPP~POWHEG prediction is shown in red while the NLO and LO
      predictions of \textsf{MCFM} appear as black stars and blue dots respectively.]{
    \includegraphics[scale=0.45]{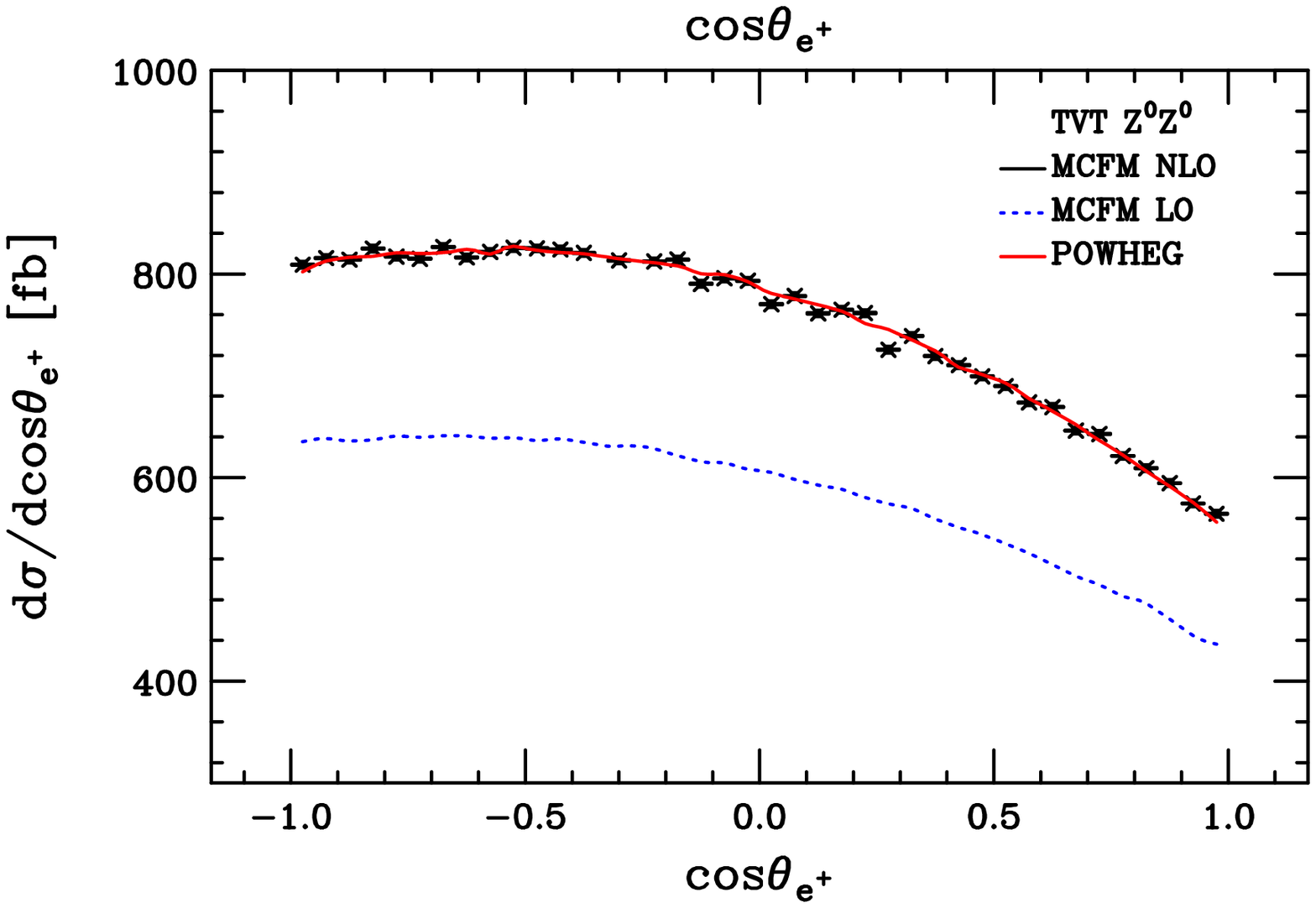}
  }%
  \hfill%
  \subfigure[Predictions for the $W^{+}$ boson
      $p_{T}$ spectrum in $W^{+}W^{-}$ pair production at the LHC, assuming
      a hadronic centre-of-mass energy of $\sqrt{s}=14$~TeV. Leading-order
      \HWPP\ results are seen in blue while those of \textsf{MC@NLO} and the 
      \HWPP~POWHEG simulation are visible as black and dashed red lines.]{
    \includegraphics[scale=0.45]{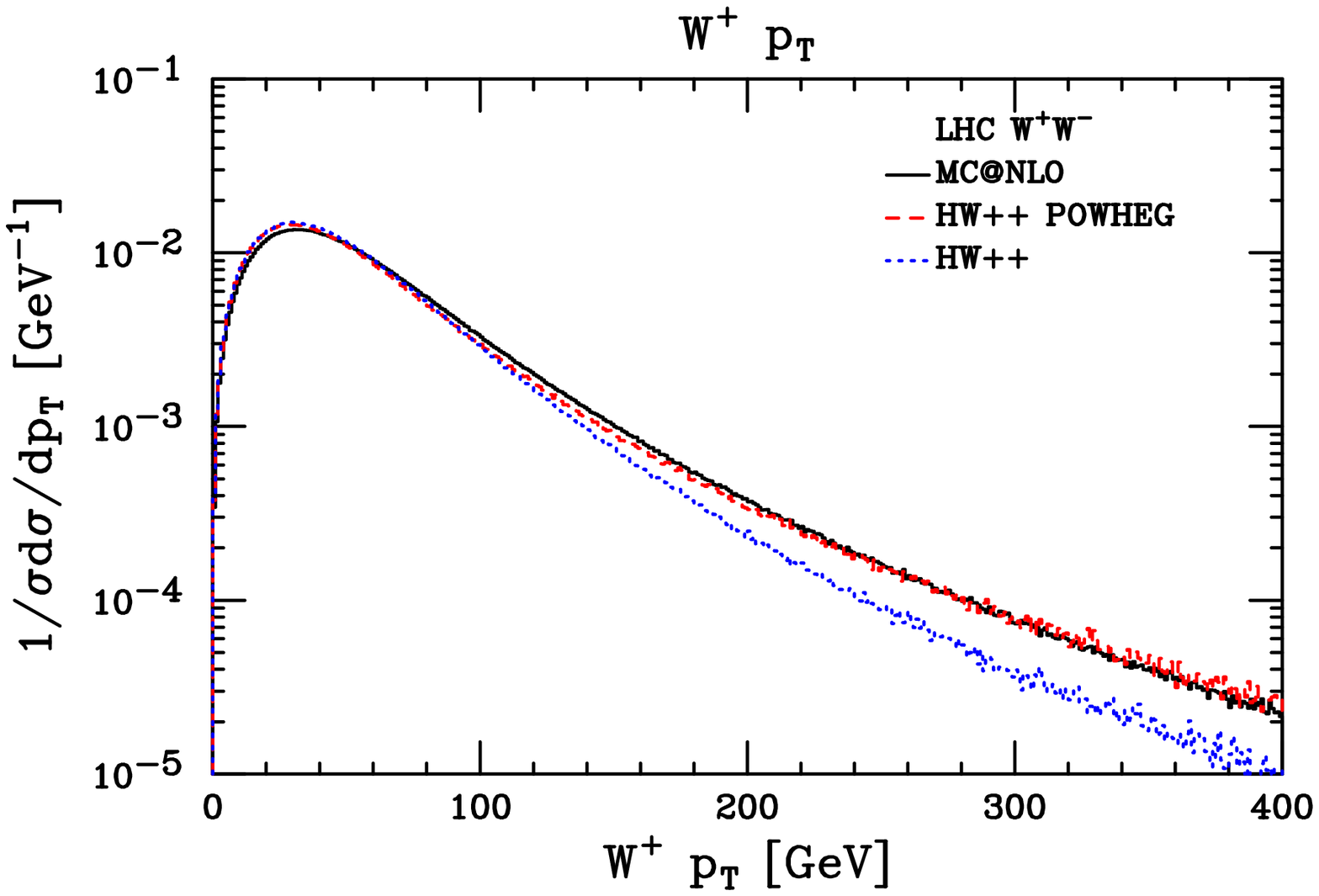}
  }

    \hfill{}
    \caption{Comparison of \HWPP, \textsf{MCFM} and \textsf{MC@NLO} for di-boson production.}
    \label{fig:dibosons}
\end{centering}
\end{figure}

  The simulation of weak boson pair production employs the NLO computations
  of Refs.~\cite{Mele:1990bq,Frixione:1992pj,Frixione:1993yp} and its
  construction has been documented in
  detail in Ref.~\cite{Hamilton:2010mb}, together with substantial comparisons
  to the \textsf{MCFM} and \textsf{MC@NLO} programs~\cite{Campbell:1999ah,Frixione:2008ym}.
  These studies show excellent agreement with \textsf{MCFM} and generally
  good agreement with \textsf{MC@NLO}. Two examples of these comparisons can be seen
  in Fig.~\ref{fig:dibosons}. Note that spin correlations in the vector
  boson decays are included at tree level for all production modes and
  decay channels. The simulation of leptonically decaying vector bosons is
  enhanced, including leading-log photon emission effects resummed according
  to the YFS formalism~\cite{Yennie:1961ad,Hamilton:2006xz}.

  The NLO matrix element class, \HWPPClass{MEPP2VVPowheg}, inherits from
  that of the leading-order matrix element, \HWPPClass{MEPP2VV}, thus 
  the selection of particular final states is performed using the 
  \HWPPParameter{MEPP2VVPowheg}{Process} interface. Any one of the five
  production modes $W^{+}W^{-}$, $W^{\pm}Z^0$, $Z^0Z^0$, $W^{+}Z^0$, and 
  $W^{-}Z^0$, can be activated by setting \HWPPParameter{MEPP2VVPowheg}{Process}
  equal to 1-5 respectively in the input files.
  Lastly we note that, as with all other \HWPP~POWHEG simulations, the
  default renormalization and factorization scale chosen for generating
  the underlying Born kinematics, {\it i.e.}~the initial $q\bar{q}\rightarrow VV$
  configuration, is given by the invariant mass of the colourless final-state
  system. In generating the hardest emission from this configuration, the
  renormalization and factorization scales used to evaluate the 
  real cross section are set to the transverse momentum of the emission, as
  mandated by the POWHEG formalism.

\subsection[$e^+e^-\to q \bar q$]{\boldmath{$e^+e^-\to q \bar q$}}

The next-to-leading order matrix element for $e^+e^-\to q \bar q$ is implemented,
including the masses of the quarks, using the POWHEG scheme
in the \HWPPClass{MEee2gZ2qqPowheg} class. 
The virtual corrections were taken from Ref.\,\cite{Jersak:1981sp}.
The real contribution to both the hardest emission and the $\bar{B}$
function was calculated using the internal helicity amplitude code.
We use the massive dipole scheme of Ref.\,\cite{Catani:2002hc} to subtract
the singularities from the real contribution for the calculation of
$\bar{B}$ and to separate the singular regions for the generation of the
hardest emission.

\begin{figure}

  \subfigure[The variation of the cross section  $e^+e^-\to b\bar b$ production at next-to-leading order.]{
\label{fig:powhegeea}
  \includegraphics[angle=90,width=0.48\textwidth]{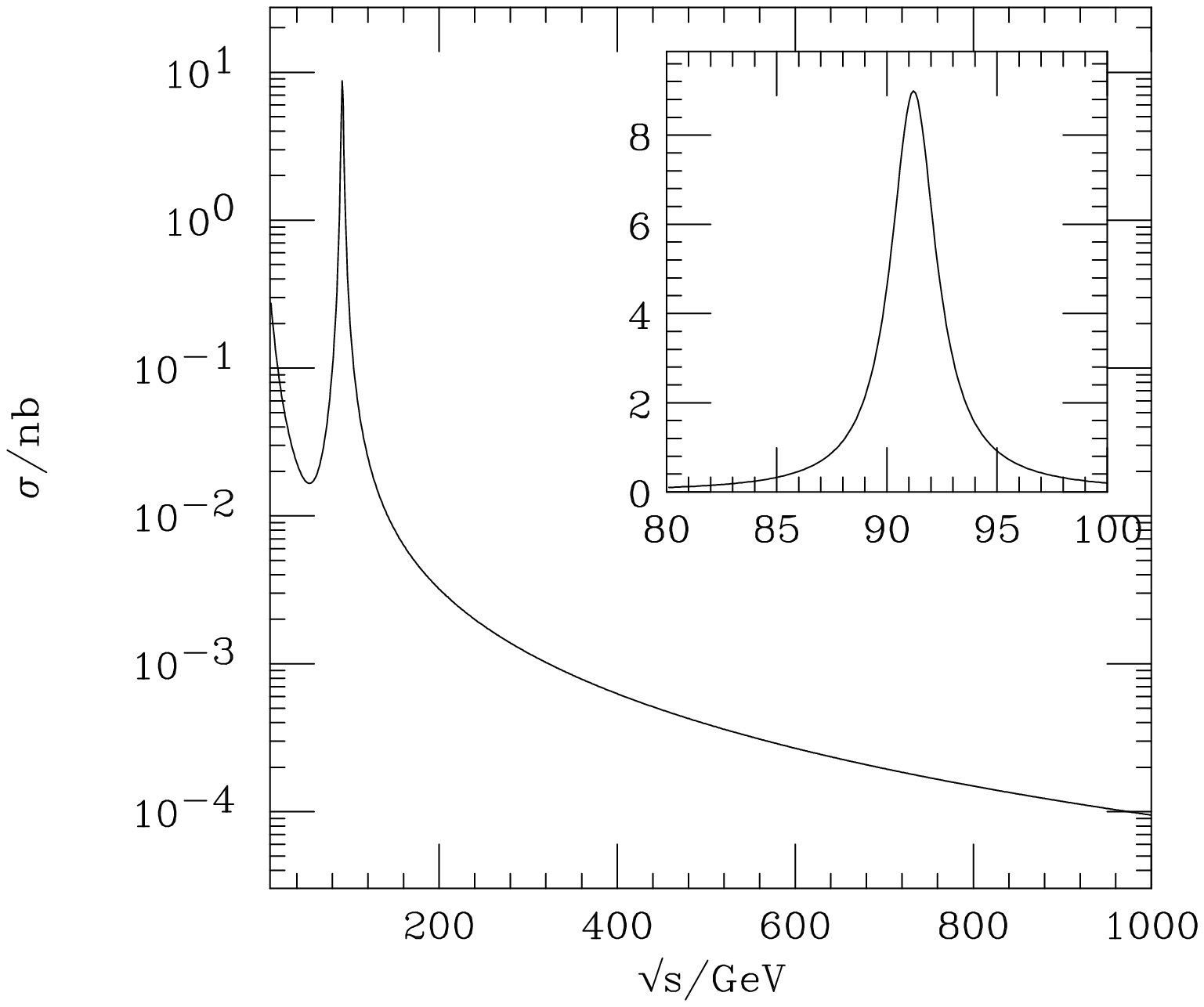}
  }%
  \hfill%
  \subfigure[The thrust distribution compared to DELPHI data~\cite{Abreu:1996na}
at 91.2\,GeV.]{%
\label{fig:powhegeeb}
  \includegraphics[angle=90,width=0.45\textwidth]{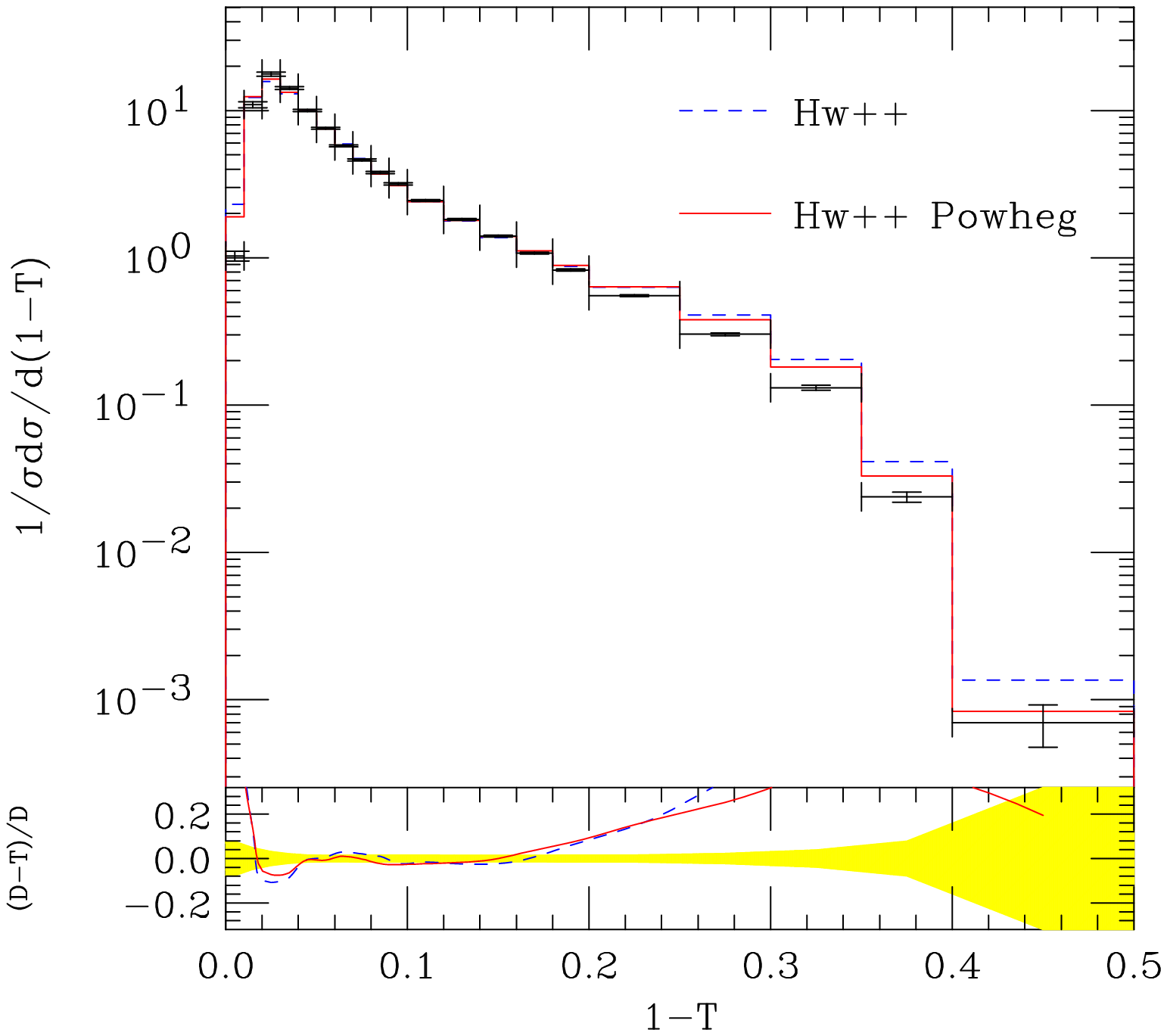}
  }
\caption{Cross section for $e^+e^-\to b\bar b$ and thrust distribution at LEP.}
\label{fig:powhegee}
\end{figure}

The results for the cross section for both bottom and top production are in
excellent agreement with the analytic results given in the appendix of 
Ref.\,\cite{Catani:2002hc}. The variation of the cross section for the
production of $b\bar b$ pairs in $e^+e^-$ collisions with the centre-of-mass
energy is shown\footnote{The analytic result is not shown as it
is indistinguishable from the \HWPP\ result.} in Fig.\,\ref{fig:powhegeea}.
The results of this new simulation are in comparable, albeit marginally
better, agreement with the data from the LEP experiments, for example
the distribution of the thrust shown in Fig.\,\ref{fig:powhegeeb}, than
the default \HWPP\ approach including a matrix element correction.

\subsection{Higgs Decay}

There are large QCD corrections to the partial widths for $h^0\to q \bar q$.
As with the simulation of $e^+e^-\to q \bar q$ we use the dipole
subtraction scheme of Ref.\,\cite{Catani:2002hc} both to subtract
the singularities from the real contribution for the calculation of
$\bar{B}$ and to separate the singular regions for the generation of the
hardest emission. The internal helicity amplitude code was used to 
calculate the real emission and the results of Ref.\,\cite{Braaten:1980yq} for 
the virtual contribution. In order to correspond as closely as possible
with the masses used in the parton shower, we use the running mass for the
coupling of the Higgs boson to the quarks, in order to resum the large corrections,
and the pole mass in the kinematics. The resulting simulation is available
in the \HWPPClass{SMHiggsFermionsPOWHEGDecayer} class.

\subsection{Summary}

  For convenience, we give the full complement of NLO-accurate POWHEG
  simulation classes included in \HWPP:
\begin{itemize}
\item the \HWPPClass{MEqq2gZ2ffPowheg} and \HWPPClass{MEqq2W2ffPowheg} 
  classes for the production and decay of the
  $\gamma^*/Z^0$ and $W^\pm$ bosons, respectively, in the Drell-Yan process;
\item the \HWPPClass{MEPP2HiggsPowheg} class for the production of the Higgs boson
  via the gluon-gluon fusion process;
\item the \HWPPClass{MEPP2ZHPowheg} and \HWPPClass{MEPP2WHPowheg}
  classes for the production of the Higgs boson
  in association with the  $Z^0$ and $W^\pm$ bosons, respectively;
\item the \HWPPClass{MEPP2VVPowheg} class for vector boson pair production
  processes;
\item the \HWPPClass{MEee2gZ2qqPowheg} class for $e^+e^-\to q \bar q$;
\item the \HWPPClass{SMHiggsFermionsPOWHEGDecayer} class for Higgs boson
      decay into quark-antiquark pairs.
\end{itemize}
  Examples illustrating the use of all of these POWHEG processes
  can be found in the {\tt LHC-Powheg.in} and {\tt TVT-Powheg.in} example
  files provided in the release.

\section{Colour Reconnection}

As of this release of \HWPP{}, a model for colour reconnection is included. It is
implemented in the \HWPPClass{ColourReconnector} class. The model can be regarded as an
extension of the cluster model \cite{Webber:1983if}, which is used for hadronization in
\HWPP{}. We would like to stress, however, that this colour reconnection model differs
from the one used in (\fortran-) \HW{}\cite{Marchesini:1991ch}, which is based on the
spacetime structure of the event as described in Ref.~\cite{Webber:1997iw}.

\subsection{Review of Cluster Hadronization}

Hadronization in \HWPP{} is based on the pre-confinement property of perturbative QCD
\cite{Amati:1979fg}. According to that, a parton shower evolving to the cut-off scale
$Q_0$ ends up in a state of colourless parton combinations with finite mass of
$\mathcal{O}(Q_0)$. In the cluster hadronization model, these parton combinations -- the
clusters -- are interpreted as highly excited pre-hadronic states. They act as a starting
point for the generation of hadrons via cluster decays, which is possibly performed in
multiple steps. This hadronization model is described in more detail in
Ref.~\cite{Bahr:2008pv}.

\subsection{Colour Reconnection Algorithm}

Colour reconnection in the cluster model occurs at the stage where clusters are formed from the
parton-shower products. Starting with the clusters that are produced generically by
virtue of pre-confinement, the cluster creation procedure is slightly
modified. This is done by allowing pairs of clusters to be `reconnected'. That means the
coloured constituent of cluster $A$ and the anti-coloured constituent of cluster $B$ form a
new cluster, as do the remaining two partons.

The following steps are performed for each cluster:
\begin{enumerate}
\item loop over all other existing clusters and choose the one where a reconnection of the two clusters would result in the
smallest sum of cluster masses;
\item if such a reconnection possibility is found, accept it with probability
\HWPPParameter{ColourReconnector}{ReconnectionProbability}.
\end{enumerate}

\subsection{Effects on Observables}

\begin{figure}[t!]
  \subfigure[Differential 2-jet rate, data from Ref.~\protect\cite{Pfeifenschneider:1999rz}.]{%
    \label{fig:leptune:a}%
    \includegraphics[width=0.47\textwidth]{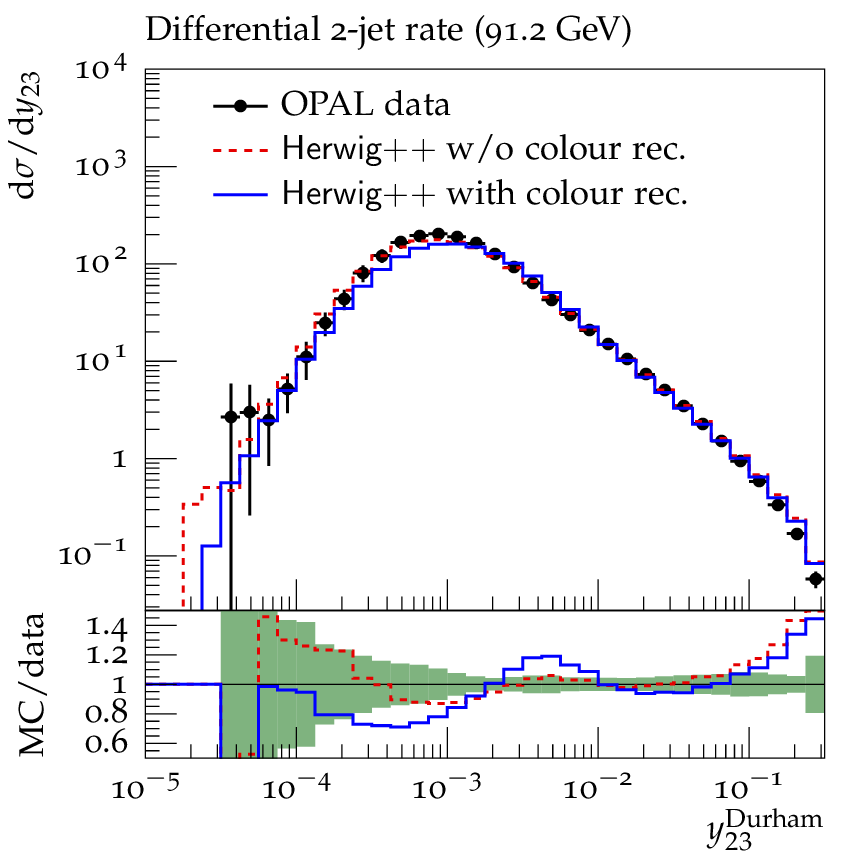}%
  }%
  \hfill%
  \subfigure[Momentum of charged particles scaled to the beam momentum, data from Ref.~\protect\cite{Abreu:1996na}.]{%
    \label{fig:leptune:b}%
    \includegraphics[width=0.47\textwidth]{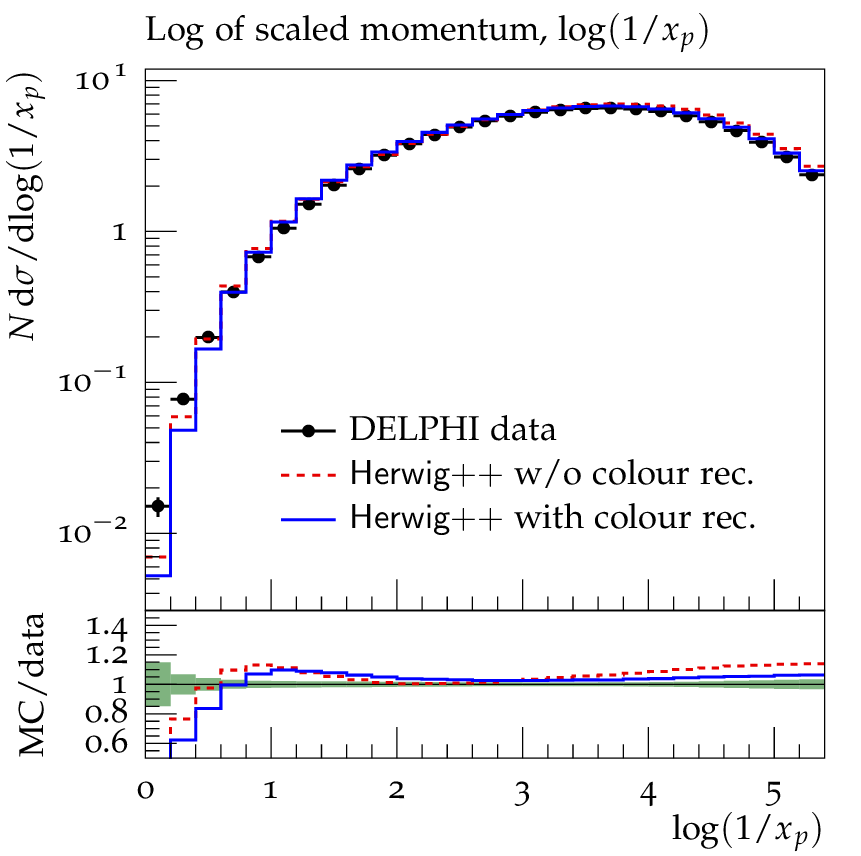}%
  }
  \caption{Example distributions comparing an old tune of \HWPP\ without colour reconnections
           to a new tune including colour reconnections.}
  \label{fig:leptune}
\end{figure}

A re-tuning of the parton-shower and hadronization-related parameters to LEP data
was inevitable since this model for
non-perturbative colour reconnection affects hadronization. We find agreement with
the LEP data at a similar level to that achieved by the old \HWPP\ tune without
colour reconnection. As an example, the 2-jet rate using the Durham jet algorithm is shown
in Fig.~\ref{fig:leptune:a}. A further illustration is given in Fig.~\ref{fig:leptune:b},
which shows the scaled momentum of charged particles at LEP.

Since colour reconnection explicitly affects the cluster mass spectrum, the average
mul\-ti\-pli\-ci\-ty of charged particles changes. This observable was measured to be
$\langle N_{\mathrm{ch}} \rangle = 20.92 \pm 0.24$ at LEP for $\sqrt{s} = 91.2$~GeV
\cite{Abreu:1996na}. With $\langle N_{\mathrm{ch}} \rangle _{\mathrm{Hw++}} = 20.73$
\HWPP{} (with colour reconnection) coincides with this measurement.

\begin{figure}
  \subfigure[Charged-particle multiplicity as a function of pseudorapidity.]{%
    \label{fig:ATLAS:a}%
    \includegraphics[width=0.48\textwidth]{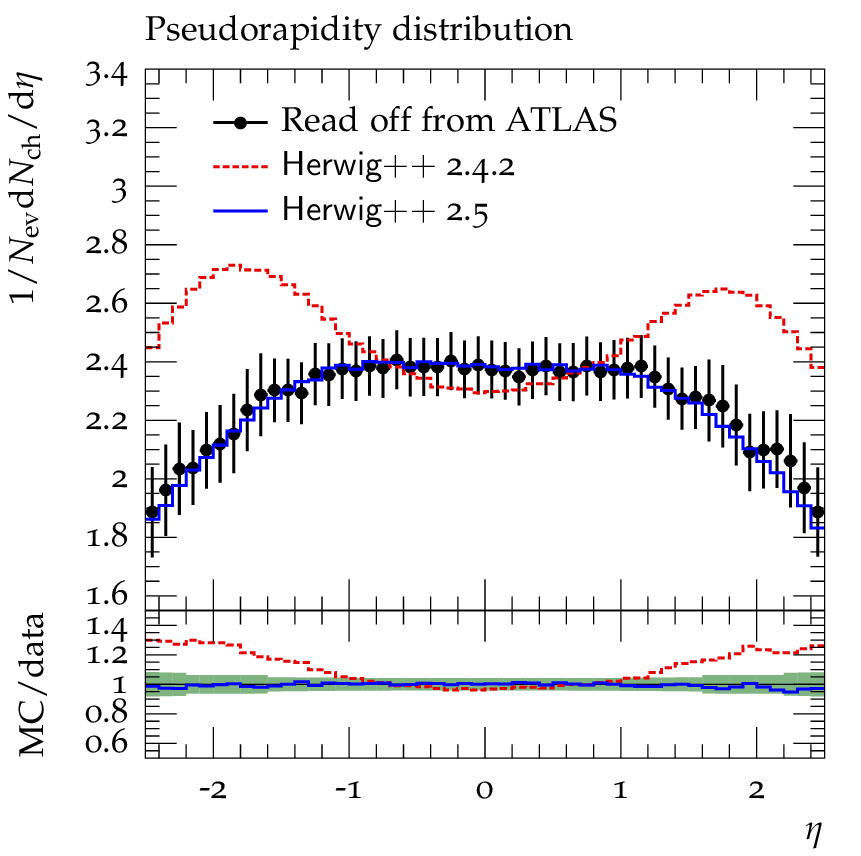}%
  }  \hfill%
  \subfigure[Average transverse momentum versus charged-particle multiplicity.]{%
    \label{fig:ATLAS:b}%
    \includegraphics[width=0.48\textwidth]{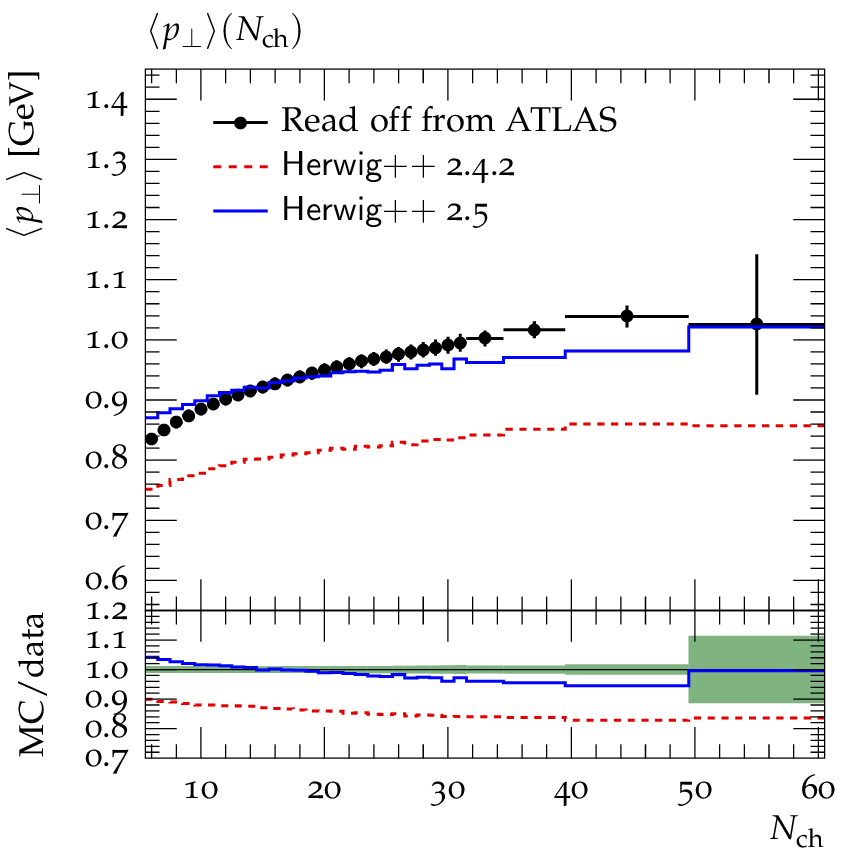}%
  }
  \caption{%
  Comparison of \HWPP{} 2.4.2 and \HWPP{} 2.5 to ATLAS minimum-bias distributions at
  $\sqrt{s}=0.9~\mathrm{TeV}$ with $N_{\mathrm{ch}} \ge 6$, $p_{\perp} > 500~\mathrm{MeV}$
  and $|\eta| < 2.5$. The ATLAS data were read off from plots published in
  Ref.~\cite{ATLAS-CONF-2010-031}.
  }
  \label{fig:ATLAS}
\end{figure}

Major improvements, however, are achieved in the description of minimum bias events and
the underlying event in hadron collisions. In Fig.~\ref{fig:ATLAS} we show the
pseudorapidity distribution of charged particles and their average transverse momentum as
function of the charged-particle multiplicity for minimum bias events at
$\sqrt{s}=0.9~\mathrm{TeV}$ at the LHC. This direct comparison of the current
\HWPP{} release to its predecessor shows the progress in modelling minimum-bias events,
partially enabled by the presence of non-perturbative colour reconnection.

A second ingredient for the latest developments in the minimum bias and underlying event
modelling is a modification of the model for soft interactions in \HWPP{} \cite{Bahr:2009ek}.
The previous model generated additional soft scatters, implemented
as soft gluon collisions in di-quark scatterings, in a way that their colour structure is
entirely disconnected from the rest of the event. We extend this choice by allowing a
colour connection between the soft scatters and the beam remnants. The parameter
\HWPPParameter{HwRemDecayer}{colourDisrupt} is the probability of a soft scatter to be
disconnected.

\HWPP{} is also capable of reproducing minimum-bias data at 7~TeV, as well as
underlying-event-related observables for both 0.9 TeV and 7 TeV. Different tunes for each
application have to be used, though, since a proper description of all energies
with the same set of parameters is not possible with the present model. A more generally
applicable model is intended for future \HWPP{} releases. 
The default parameters in the current release remain those used in the previous version,
{\it i.e.}~no colour reconnections are included. The best parameters including
colour reconnections for different processes and energies can be found at 
\begin{quote}\tt
  \href{http://projects.hepforge.org/herwig/trac/wiki/MB\_UE\_tunes}
       {http://projects.hepforge.org/herwig/trac/wiki/MB\_UE\_tunes}
\end{quote}

\section{Diffractive and Photon-Initiated Processes}

Some interesting forward processes in hadron--hadron collisions have been added
in this version of \HWPP, namely photon-initiated processes and inclusive hard diffractive processes.

\subsection{Photon-Initiated Processes}

\begin{figure}
\centering
\includegraphics[scale=0.3]{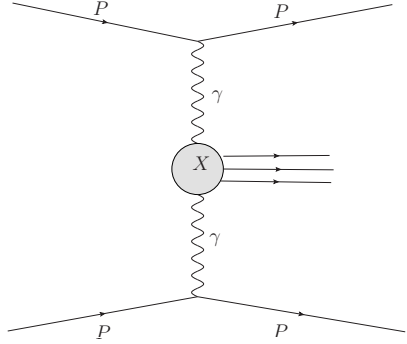}
\caption{Diagram of a two-photon induced process.}
\label{fig:gamma-gamma}
\end{figure}

In photon-initiated processes photons are emitted by the incoming protons and their interaction yields 
a system $X$ which is separated by large rapidity gaps in the forward region from the outgoing scattered protons.
The protons remain intact in the interaction and they are only scattered by a small angle. 
A sketch of the interaction $pp \rightarrow p + \gamma + \gamma+p \rightarrow  p \oplus X \oplus p$, 
where $\oplus$ stands for a rapidity gap, can be seen in Fig.~\ref{fig:gamma-gamma}. 
The emission of photons by protons is well described by Quantum Electrodynamics in the framework 
of Equivalent Photon Approximation \cite{Budnev:1974}. The Photons are almost real~(the
photon virtuality $Q^2 \sim 0$) and therefore 
the cross section can be factorized into the photon fluxes and the sub-matrix element. 
The photon flux is implemented in the \ThePEGClass{BudnevPDF} class. The photons are generated according to
\begin{equation}
 dN = \frac{\alpha}{\pi}\frac{dE_{\gamma}}{E_{\gamma}}\frac{dQ^2}{Q^2}
\left[(1-\frac{E_{\gamma}}{E})(1-\frac{Q_{\min}^{2}}{Q^{2}})F_{E} + \frac{E_{\gamma}^2}{2E^2}F_M \right],
\end{equation}
where $E_{\gamma}$ is the photon energy, $E$ is the energy of proton and $Q_{\min}^2$ is the minimum virtuality of the photon 
allowed by the kinematics. The electric and magnetic form factors, $F_E$ and $F_M$, are given by
\begin{subequations}
\begin{eqnarray}
F_E &=& (4m_{p}^2G_{E}^2 + Q^{2}F_M)/(4m_{p}^2 + Q^2), \\
G_{E}^2 &=& F_M/\mu_{p}^{2} = (1+Q^2/Q_{0}^{2})^{-4},
\end{eqnarray}
\end{subequations}
with the magnetic moment of the proton $\mu_{p} =7.78$, the fitted scale $Q^{2}_{0} =
0.71$~GeV$^2$ and the proton mass $m_p$.  An example input file example for LHC settings,
\texttt{LHC-GammaGamma.in}, is provided in the release.

\subsection{Diffractive Processes}

\begin{figure}
\centering
\includegraphics[scale=0.3]{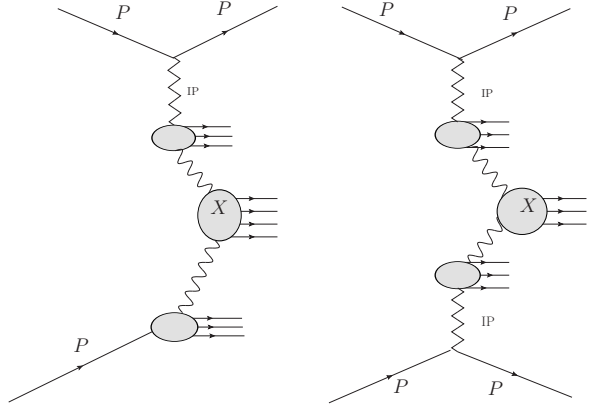}
\caption{Examples of single diffractive (left) and double pomeron exchange (right) processes.}
\label{fig:sd-ex}
\end{figure}

In the double pomeron exchange process both protons are left intact by the scattering process
and two rapidity gaps occur in forward regions. In single diffractive processes one proton dissociates, while the other is left intact,
and a rapidity gap appears on one side of the event. Examples of both processes are shown in Fig.~\ref{fig:sd-ex}.
The Ingelman-Schlein model~\cite{Ingelman:1984ns} has been implemented to describe those
processes. In this model the cross section is factorizes into 
a diffractive distribution function and the cross section of the sub-process, $\sigma = f_{D}(x, Q^2, x_\mathbb{P}, t) \otimes \sigma_{\rm sub}(x, Q^2)$. 
The diffractive distribution function can be further decomposed into the pomeron/reggeon flux and distribution functions, respectively,
\begin{equation}
 f_{D}(x, Q^2, x_\mathbb{P}, t)=f_{\mathbb{P}/p}(t,x_\mathbb{P})f_{i/\mathbb{P}}(Q^{2},x)+\eta_{\mathbb{R}}f_{\mathbb{R}/P}(t,x_{\mathbb{R}})f_{i/\mathbb{R}}(Q^{2},x),
\end{equation}
where $f_{\mathbb{P}}(t,x_{\mathbb{P}})$ and $f_{\mathbb{R}}(t,x_{\mathbb{R}})$ are pomeron and reggeon fluxes.
We use the Regge theory  inspired form
\begin{equation}
 f_{\mathbb{P}(\mathbb{R})}(t,x_{\mathbb{P}(\mathbb{R})})=A_{\mathbb{P}(\mathbb{R})}e^{\beta_{\mathbb{P}(\mathbb{R})}t}/x^{2(\alpha_{\mathbb{P}(\mathbb{R})0} - \alpha_{\mathbb{P}(\mathbb{R})}'t)-1},
\end{equation}
 where $\alpha_{\mathbb{P}(\mathbb{R})0}$ and $\alpha_{\mathbb{P}(\mathbb{R})}'$ are the pomeron/reggeon intercept and slope, respectively. 
 Both the pomeron and reggeon fluxes are implemented in the \HWPPClass{PomeronFlux} class because they differ 
 only in choice of  $\alpha_0$ and $\alpha'$. The pomeron distribution function $f_{i/\mathbb{P}}(Q^{2},x)$ is 
 implemented in the \HWPPClass{PomeronPDF} class. \HWPP\ includes the 2006 Fit A, 2006 Fit B and 2007 fits of 
 the pomeron distribution function measured at HERA~\cite{Hera:2006,Hera:2007}. The \HWPPClass{PomeronPDF} class allows the user to 
 freeze or extrapolate the pomeron distribution functions outside the limits of the experimental fit. In the case of the reggeon 
 distribution function, $f_{i/\mathbb{R}}(Q^{2},x)$, an external~(pion) parton distribution function from \textsf{LHAPDF}~\cite{Whalley:2005nh}
  can be used and \HWPP\ provides only the interface class 
 \HWPPClass{ReggeonPDF} which mimics a reggeon beam particle and calls an external user-defined PDF. 
 Interfaces to change parameters such as $\alpha_{\mathbb{P}(\mathbb{R})0}$, 
 $\alpha_{\mathbb{P}(\mathbb{R})}'$ and $\eta_{\mathbb{R}}$  are provided.
 By default they are set to the values determined in the used PDF fit.

 We also provide two options for the internal valence structure of the pomeron. The 
 pomeron can be composed of either valence $q\bar{q}$ or valence gluons, the latter of which is the default setting.
 The reggeon is treated as a $q\bar{q}$ object since using the pion PDF is a usual and
 consistent approach to describe the reggeon.

 The pomeron or reggeon contributions can 
 be simulated either separately or as a mixture. An example of the settings for the LHC can be found in the \texttt{LHC-Diffractive.in} input file
 supplied with the release.
 It should be noted that the model does not include the gap survival probability factors which need to be taken into 
 account in hadron-hadron collisions. 

\section{BSM Physics}

  In addition to including more models of physics
  beyond the Standard Model~(BSM), this release includes a number
  of other improvements to the simulation of BSM physics in \HWPP.

\subsection{Process Generation}

  The inheritance structure of the code
  for simulating general or resonant $2\to2$ scattering
  processes has been changed. A new \HWPPClass{HardProcessConstructor}
  base class has been added. The existing code
  for general $2\to2$ scattering processes has been renamed 
  \HWPPClass{TwoToTwoProcessConstructor} and now inherits
  from the new \HWPPClass{HardProcessConstructor}
  base class, as does the \HWPPClass{ResonantProcessConstructor}
  class for resonant $2\to2$ scattering processes. As part
  of this change, the specific interfaces to 
  the objects needed to simulate $2\to2$ scattering processes
  have been removed from the \HWPPClass{ModelGenerator} class
  and replaced by a list of \HWPPParameter{ModelGenerator}{HardProcessConstructors}
  in order to make adding further types of hard processes
  easier.

  These changes have allowed the generation of hard
  scattering processes in BSM models
  to be extended to include some $2\to3$ processes
  involving neutral Higgs bosons.
  The production
  of Higgs bosons in association with 
  an electroweak vector boson, $W^\pm$ and $Z^0$,
  is generated using the \HWPPClass{HiggsVectorBosonProcessConstructor}
  class. The production of a Higgs boson in
  association with a heavy quark-antiquark pair is
  calculated using the \HWPPClass{QQHiggsProcessConstructor} class.
  The production of the Higgs boson via the
  vector boson fusion~(VBF) process is simulated using the
  \HWPPClass{HiggsVBFProcessConstructor} class.

  In addition, a number of improvements have been made to the hard process selection for
  $2\to2$ scattering processes in BSM models.
  The \HWPPParameter{TwoToTwoConstructor}{Processes}
  interface can be used to select between: 
  \HWPPParameterValueB{TwoToTwoConstructor}{Processes}{SingleParticleInclusive} where
  at least one of the particles in the list of \HWPPParameter{TwoToTwoConstructor}{Outgoing}
  particles must be produced, which is the previous behaviour and the new default;
  the \HWPPParameterValueB{TwoToTwoConstructor}{Processes}{TwoParticleInclusive}
  option where both the particles produced must be in the list
  of \HWPPParameter{TwoToTwoConstructor}{Outgoing} particles; and the 
  \HWPPParameterValueB{TwoToTwoConstructor}{Processes}{Exclusive} option where 
  only two particles are allowed in
  the list of \HWPPParameter{TwoToTwoConstructor}{Outgoing} particles and both of these must be produced.
  In addition, the \HWPPParameter{TwoToTwoConstructor}{Excluded} interface can be used
  to forbid specific particles as intermediate particles in the scattering process.
  Similarly the \HWPPParameter{TwoToTwoConstructor}{ExcludedVertices} interface can be used to
  forbid specific vertices in the hard scattering processes

  The scale choice in the hard $2\to2$ scattering processes in BSM models can now be changed
  via the new \HWPPParameter{TwoToTwoConstructor}{ScaleChoice} interface to choose 
  between $\hat s$~(default for colour-neutral intermediates) and the transverse 
  mass~(default for all other processes).

\subsection{ADD Model}

  In the ADD Model~\cite{ArkaniHamed:1998rs,ArkaniHamed:1998nn,Lykken:1996fj,Witten:1996mz,Horava:1995qa,Horava:1996ma,Antoniadis:1990ew} 
  gravity propagates in extra spatial dimensions, which have a flat metric.
  The large size of these extra dimensions leads to a tower of Kaluza-Klein excitations
  of the graviton. They can either contribute as virtual particles to Standard Model~(SM)
  processes, or be produced leading to missing energy signatures. This model is
 implemented using the conventions of Ref.\,\cite{Giudice:1998ck}.

\begin{figure}
  \subfigure[Photon + missing transverse energy cross section.]{
 \includegraphics[angle=90, width=0.465\textwidth]{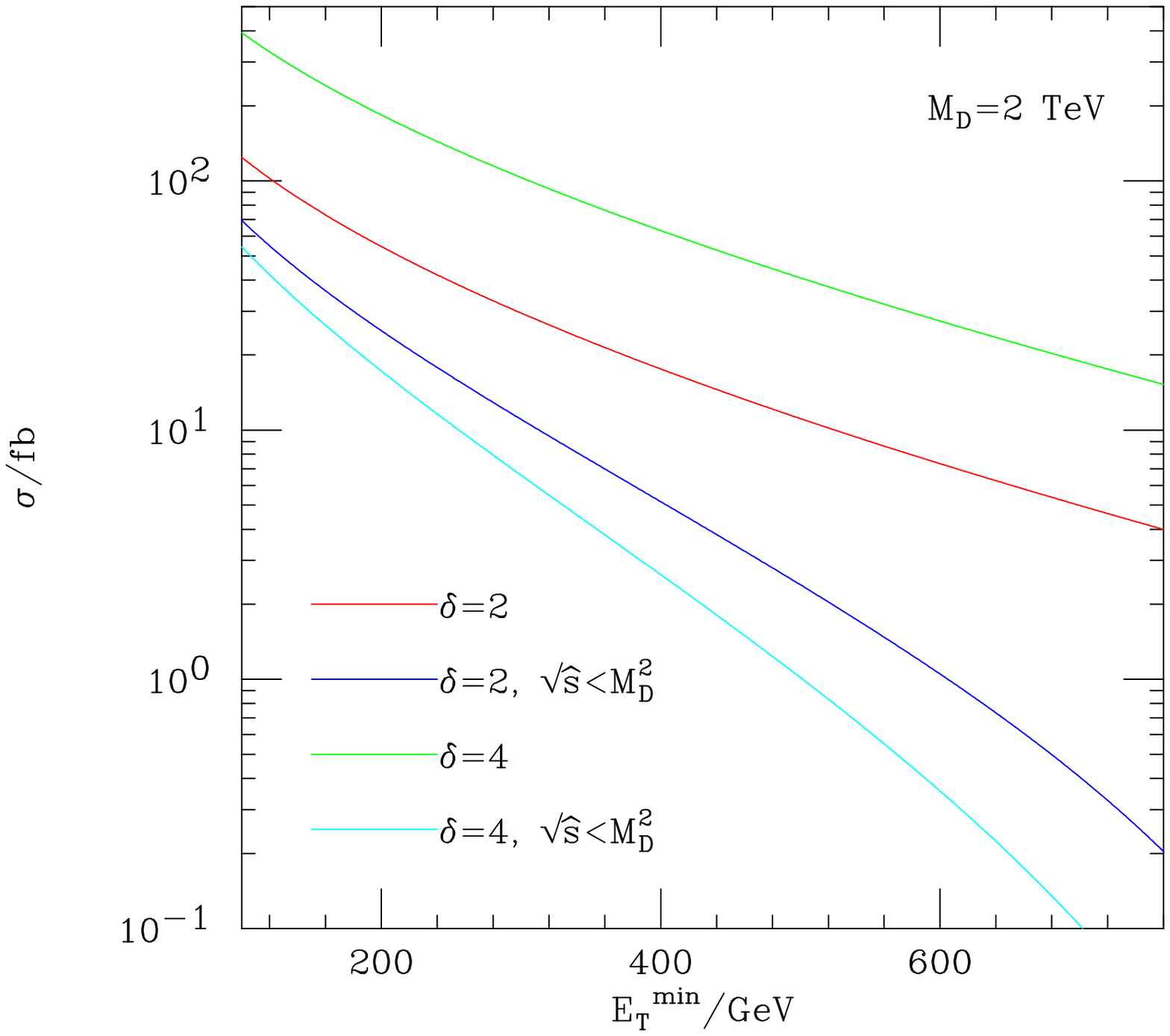}
  }  \hfill%
  \subfigure[Jet + missing transverse energy cross section.]{
\includegraphics[angle=90, width=0.48\textwidth]{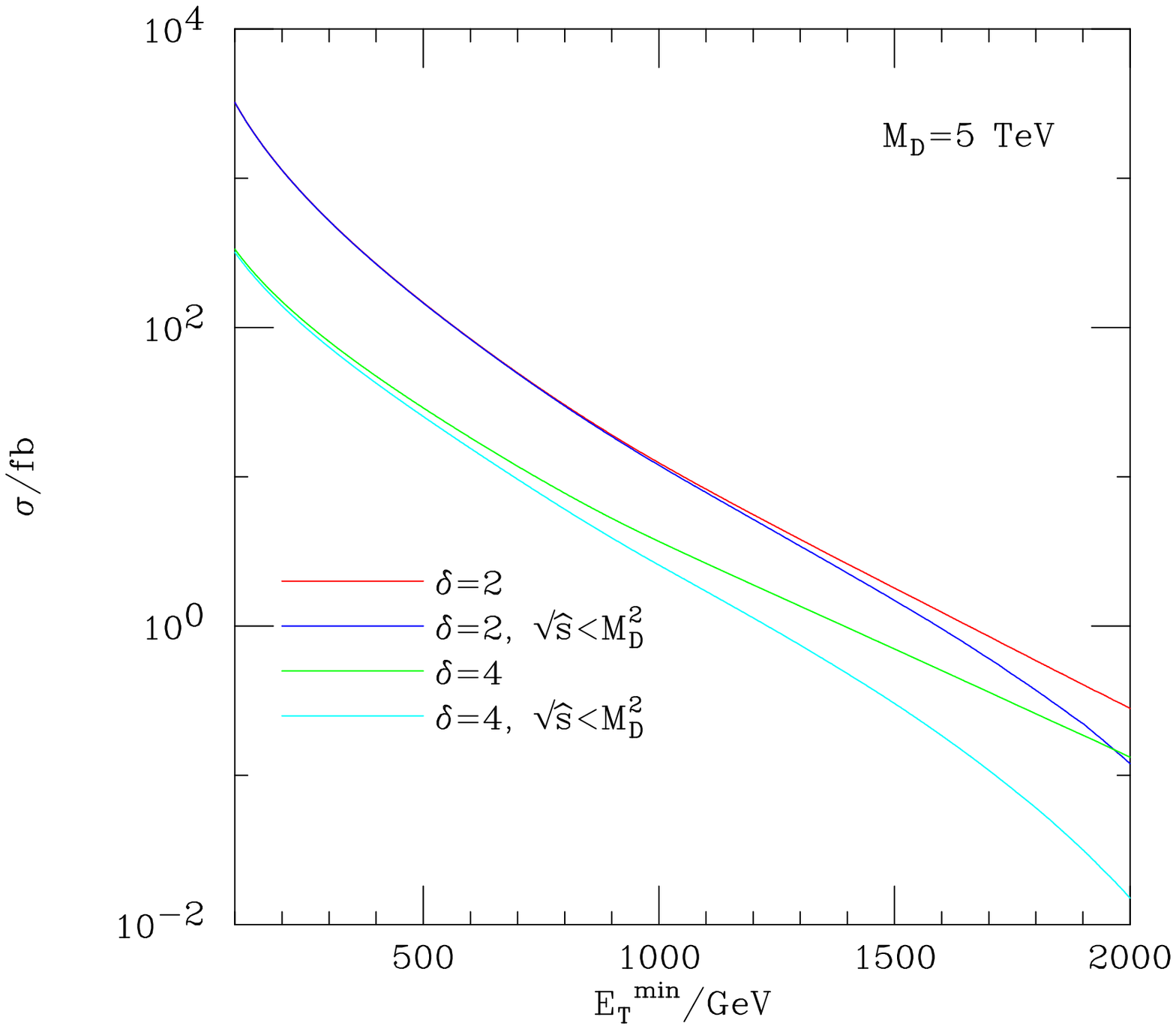}
  }
\caption{The total photon and jet + missing transverse energy cross section at the LHC. The photons/jets are required to have  transverse energy greater than $E_T^{\rm min}$.
The pseudorapidity of the photons/jets are required to satisfy, $|\eta_\gamma|<2.5$ and
$|\eta_{\rm jet}|<3.0$, respectively. The cross section is integrated over either all values of $\hat s$ or $\hat s<M^2_D$. This figure is based on Figs.\,3~and~5 of Ref.\,\cite{Giudice:1998ck}.\vspace*{-0.6em}}
\label{fig:add}
\end{figure}
  An example input file for the LHC, \texttt{LHC-ADD.in}, is provided, and an example of the total
 cross section for photon or jet production with missing transverse energy is
 shown in Fig.\,\ref{fig:add}.

\subsection{Leptoquarks}
\vspace*{-0.4em}

Fermion masses may arise from the mixing of elementary fermions with
composite, fermionic resonances of a strong
sector~\cite{Kaplan:1991dc} responsible for the breaking of the
$SU(2)_L \times U(1)_Y$ electroweak symmetry of the SM. It follows that
this strongly-coupled sector must also be charged under colour $SU(3)$
and must contain, at the very least, colour-triplet fermionic
resonances that can mix with the elementary colour triplets to make
the observed quarks. It is reasonable to expect that such a
strongly-coupled sector will contain other coloured resonances. These
may be bosonic and, depending on their charges, may couple to a quark
and a lepton. These leptoquark resonances may be light if they arise
as pseudo-Nambu Goldstone bosons and make an ideal target for LHC
searches. They will decay exclusively to third-generation fermions due
to suppression of the couplings to light fermions.

The present implementation includes non-derivatively coupled leptoquarks, that couple to Standard Model
fermions as in Eq.~(2.4) of Ref.~\cite{Gripaios:2010hv} and single-derivatively coupled
leptoquarks such as those in Eq.~(2.5) of Ref.~\cite{Gripaios:2010hv}. In the case of the
derivatively coupled leptoquarks, the simplification that the
primed lepton and primed quark couplings are equal has been made. 

\vspace*{-0.6em}
\subsection{NMSSM}
\vspace*{-0.4em}

The Next-to-Minimal Supersymmetric Standard Model~(NMSSM) extends the Minimal Supersymmetric Standard Model~(MSSM)
with the addition of a singlet Higgs superfield, $\hat S$. This leads to a larger particle content than the MSSM,
{\it i.e.}~3 scalar Higgs bosons, 2 pseudoscalar Higgs bosons and 5 neutralinos. 
This model has been shown to overcome or reduce many problems associated with the MSSM.
For recent reviews of the NMSSM see Refs.~\cite{Ellwanger:2009dp,Maniatis:2009re}.

The phenomenologically relevant trilinear interactions of the Higgs
bosons with fermions, sfermions, gauginos, gauge bosons, gluons, photons
and the other Higgs bosons are included. All the other
important interactions are inherited from the implementation of
the Minimal Supersymmetric Standard Model. More details are available in Ref.~\cite{NMSSM}.

\subsection{Transplanckian Scattering Model}

The interaction described by the Transplanckian matrix element is
$2\to2$ high-energy gravitational scattering of partons using the eikonal
approximation~\cite{Giudice:2001ce}. The approximation is valid in the high-energy, low-angle
scattering regime, where the centre-of-mass scattering angle of the
incoming parton $\hat{\theta} \rightarrow 0$ or, in terms of
Mandelstam variables, $-\hat{t}/\hat{s} \rightarrow 0$. The
implementation allows variation of the Planck scale, $M_D$, as well as
the number of extra dimensions, up to a maximum of $6$. 

\subsection{Minor Changes to BSM Models}

  The \HWPPClass{SusyBase} class has been modified so that it can
  read the SLHA parameter file from the header of a Les Houches event~(LHE)
  file. The name of the LHE file must be given via the \texttt{setup}
  command to the model class instead of the SLHA file.  

  The vertices for the interactions of gravitinos in SUSY models
  have been added to allow the decay of the next-to-lightest 
  supersymmetric particles~(NLSP) in gauge-mediated SUSY
  breaking models. In addition, the approach used to calculate
  Rarita-Schwinger spinors for spin-$\frac32$ particles
  has been changed to improve the numerical stability
  for very light spin-$\frac32$ particles, such as the 
  gravitino.

  Similarly, the vertex for the flavour-changing interaction of
  the stop quark, neutralino and charm quark has been added to 
  allow the simulation of the decay of the lightest stop at
  parameter points where its mass is lighter than that of the
  lightest chargino.

  In general, \HWPP\ does not include
  some three- and four-point vertices which do not have any 
  phenomenologically relevant collider signals, for
  example three- and four-point self couplings of the Higgs
  boson and the four-point coupling of two vector and two
  Higgs bosons. Although these interactions are not relevant
  for the processes we include in the Standard Model, they can be
  in BSM models, due to the additional particle content. 
  We have therefore extended the implementation of the Standard Model to
  include the triple Higgs boson self coupling and the coupling
  of two Higgs bosons to two vector bosons to make implementing these
  vertices in BSM models simpler. These vertices are also implemented
  in both the MSSM and NMSSM.

  The handling of the colour flows in $2\to2$ hard scatterings 
  has been changed 
  to improve the generality of the code and make future
  extensions easier. There should be no changes to the results
  for the currently implemented models. In addition, a number
  of improvements were made for the colour flows involving
  colour singlet exchange in the $s$-channel to allow
  the simulation of resonant Higgs boson production.

  The \textsf{Helicity} classes in \textsf{ThePEG} have been significantly cleaned
  up, leading to major simplification of the vertex
  implementations for all Standard Model and BSM physics models in \HWPP. The
  separate \ThePEGClass{SpinBase} and \ThePEGClass{SpinInfo}
  classes have been merged into the new  \ThePEGClass{SpinInfo} class significantly
  simplifying the code structure and reducing the number of pointer casts.
  These changes, which should be transparent to most
  users, have been made to the helicity amplitude code in order
  to improve the structure of the code and make future extensions
  easier.

\section{New Matrix Elements}

  A number of new matrix elements are included in this release:
\begin{itemize}
\item the \HWPPClass{MEPP2HiggsVBF} for the production of the Higgs
  boson via electroweak vector boson fusion in hadron--hadron
  collisions\footnote{This matrix element is based on the dominant
    $t$-channel contribution to this process.};
\item the \HWPPClass{MEPP2VV} class for the production of pairs of electroweak gauge bosons,
      $W^+W^-$, $W^\pm Z^0$ and $Z^0Z^0$ in hadron--hadron collisions;
\item the \HWPPClass{MEPP2VGamma} class for the production of an electroweak gauge boson, $W^\pm$ or $Z^0$,
      in association with a photon;
\item the \HWPPClass{MEPP2QQHiggs} class for the production of the Higgs boson in association with a 
      $t\bar t$ or $b\bar b$ pair;
\item the \HWPPClass{MEee2VV} class for the production of pairs of electroweak gauge bosons,
      $W^+W^-$ and $Z^0Z^0$ in lepton--lepton collisions;
\item the \HWPPClass{MEGammaGamma2ff} class for the production of fermion-antifermion pairs in
      photon--photon initiated processes;
\item the \HWPPClass{MEGammaGamma2WW} class for the production of $W^+W^-$ pairs in 
      photon--photon initiated processes;
\item the \HWPPClass{MEGammaP2Jets} class for the production of jets in photon--hadron collisions
      via the partonic processes $q\gamma\to q g$, $\bar{q}\gamma\to \bar{q} g$ and $g
      \gamma\to q \bar{q}$.
\end{itemize}

\section{Other Changes}

A number of changes have been made to the implementation of the 
\textsf{SplittingFunctions} in the parton shower module. Previously
a splitting function class inheriting \HWPPClass{SplittingFunction} had to be
implemented for each combination of the colours and spins of the interacting
particles. This has been changed so that now all the possible colour states
of the interacting particles are implemented in the base class while the 
inheriting classes implement the spin structure only. This makes adding
new interactions in BSM models simpler. In addition, the option of
deleting specific types of branching in the shower has been added.

A number of other more minor changes have been made.
The following changes have been made to improve the physics 
simulation:
\begin{itemize}
\item Soft QED radiation in $Z^0$ decays is now fully 1-loop by default. As
      part of this change, the old \HWPPClass{WZDecayer} class, which handled
      both $W^\pm$ and $Z^0$ decays, has been split into separate
      \HWPPClass{WDecayer} and \HWPPClass{ZDecayer} classes handling $W^\pm$ and $Z^0$
      decays, respectively.
\item Problems with numerical instabilities in the boosts applied to tau lepton
      decays have been fixed by postponing the tau decays until after the parton shower.
      A new interface setting \HWPPParameter{DecayHandler}{Excluded} in the
      \HWPPClass{DecayHandler} class
      is available to prevent decays in the shower step. By default this is only enabled
      for tau leptons.
\item The default PDF set is now the LO* set of Ref.\,\cite{Sherstnev:2007nd}.
      In addition, the PDFs used in
      the shower can now be set separately to those in the hard process using the
      \HWPPParameter{PDF}{ShowerHandler} interface of the \HWPPClass{ShowerHandler} class.
\item The shower, hadronization and underlying event parameters were retuned
      against LEP and Tevatron data respectively.
\item The mixing of $B^0_d-\bar{B}_d^0$ and $B^0_s-\bar{B}_s^0$ mesons
      has been added. The oscillation of the mesons is simulated including the
      CP-violating terms. However, there is no special treatment of those modes where both
      the meson and its antiparticle decay into the same final state.
\item The missing colour structures required for a number of decays in BSM models have been added.
\item QED radiation is now enabled in all perturbative Standard Model and
      BSM decays and all perturbative decays by default,
      where there are no strongly interacting particles involved in the decay.
\item Spin correlations are now switched on by default for all perturbative decays.
\item New interfaces to the \textsf{AcerDet}~\cite{RichterWas:2002ch} and 
      \textsf{PGS}~\cite{PGS} fast detector simulations 
      are now available in the \texttt{Contrib} directory.
\item \textsf{FastJet} \cite{Cacciari:2006sm} is now the only supported jet finder code.
  All analyses have been converted to use \textsf{FastJet}.
\item Improvements have been made to the momentum reshuffling in
       Deep Inelastic Scattering.
\item The \ThePEGClass{LesHouchesReader} class has been modified to allow 
      processes that violate baryon number to be read from LHE files.
\item A number of changes have been made in the parton shower to allow the violation of baryon number in the hard process.
      The showering of baryon number violating decays was already supported.
\item Code for the simulation of the production of ${W'}^\pm$ bosons using the POWHEG approach is
      now included in the \texttt{Contrib} directory.
\item $\mu^-$, $\nu_\mu$ and $\nu_e$, and their antiparticles, are now available as beam particles. They
      are all supported in the DIS matrix elements. $\mu^+\mu^-$ collisions are supported in the general
      matrix element code for BSM models but not yet in the hard-coded matrix elements for lepton-lepton
      scattering.
\item The option of bottom and charm quarks is now supported for heavy quark production
      in the \HWPPClass{MEHeavyQuark} class.
\item The polarization of tau leptons can now be forced in the \HWPPClass{TauDecayer}
      class to assist in studies of tau polarization.
\end{itemize}
A number of technical changes have been made:
\begin{itemize}
\item The \texttt{Tests} directory has been added. It contains many additional input files to perform more detailed tests of
      the program than are performed by default during \texttt{make check}. 
      These tests use both our own internal
      analyses and many of the analyses available in \textsf{Rivet}~\cite{Buckley:2010ar}.
\item The \LaTeX\ output has been updated. After each run, a \LaTeX\ file is produced that contains the full 
      list of citations. Please include the relevant ones in any publication.
\item A number of improvements for OS X systems have been made including fixes for the Snow Leopard release.
\item \texttt{Makefile-UserModules} now includes the \HWPP{} version number.
      In addition, the compiler flag \texttt{-pedantic} is no longer enabled
      so that user code using \textsf{ROOT} will compile.
\item The obsolete \textsf{KtJet} and \textsf{CLHEP} interfaces have been removed.
\item The zero-momentum interacting particle used for bookkeeping in Min-Bias events
         is now labelled as a pomeron. 
\item $K^0/\bar K^0$ oscillations into $K^0_{S,L}$ now occur at the production
      vertex of the kaon to give the correct decay length.
\item The default scale choice in POWHEG processes 
      is now the mass of the colour-singlet system.
\item A new \texttt{ZERO} variable has been introduced, which can be used to set
      any dimensionful quantity to zero avoiding explicit constructs like 
      \texttt{0.0*GeV}.
\item A number of improvements have been made to the implementation of three-body decays in BSM models.
\item The option of redirecting all output to \texttt{stdout} is now supported.
      The files previously ending in \texttt{-UE.out} and \texttt{-BSMinfo.out} are now appended
      to the \texttt{log} file. They now also obey the \ThePEGParameter{EventGenerator}{UseStdout} flag.
\item The detection of \textsf{FastJet} in the configuration process has been improved.
\item The interfaces to \textsf{Rivet} and \textsf{HepMC} have been moved from \HWPP\ to \ThePEG.
\item The configuration process now looks for \ThePEG{} in the location specified by \texttt{--prefix}.
\item Important configuration information is now listed at the end of the configuration process and
      in the file \texttt{config.thepeg}. Please provide this file in any bug reports.
\item The \textsf{Exception} specifiers in the definition of the \textsf{doinit()} etc.~member
      functions have been removed. This may require the removal of the exception specifier after the function name
      in some user code.
\item The SLHA EXTPAR block can now be used to set $\tan\beta$.
\item The warning threshold for branching ratios not summing to 1 has been relaxed. It is now a user interface parameter.
\item The cross section for inelastic scattering for minimum bias processes is now available in the
      \texttt{-UE.out} files.
\item A number of classes have been renamed so that tilde is correctly spelt.
\item Some deprecated interfaces in the \HWPPClass{MPIHandler} class have been removed.
\item The unused doubly heavy baryons have been deleted from the input files.
\item A number of minor changes have been made to improve the stability of the phase-space integration
      in particle decays.
\item The handling of the debugging flags has been made more consistent. The user now needs to 
      add, for example, \texttt{set LHCGenerator:DebugLevel 1} in the input file or the
      \texttt{-d 1} option on the command line in order for any debugging printout, including printing
      of events to the log file, to be enabled.
\item The handling of the PDG codes for particles has been changed in order to 
      prevent problems on 64-bit systems mapping \texttt{unsigned int}
      to \textsf{long}.
\item The version of \textsf{libtool} has been updated to 2.4. We also
      use the silent rules available in recent versions of
      \textsf{automake}, reducing the output during the compilation of \HWPP.
\item The version of \textsf{LoopTools} used in \HWPP\ has been updated to 2.6.
\item A number of \texttt{.icc} files have been removed and the corresponding code
      moved to either the corresponding \texttt{.h} or \texttt{.cc} files. Similarly
      a number of unnecessary \texttt{inline} directives and unused header files
      have been removed.
\item The implementation of the \HWPPClass{HerwigRun} class has been simplified
      by using more features of the \ThePEGClass{EventGenerator} class of \ThePEG.
\item A number of changes have been made to fix warning messages from the \textsf{Intel}
      compiler.
\item The inheritance structure of the \HWPPClass{SMHiggsWidthGenerator} has been improved
      to bring it into line with other \textsf{WidthGenerator} classes.
\item The option of applying a sequential longitudinal and then transverse boost
      in the \HWPPClass{QTildeReconstructor} in order to conserve energy and momentum,
      rather than one single boost, has been introduced primilarly for
      use by the \textsf{MC@NLO} program.
\item The interface to the \textsf{LHAPDF} package~\cite{Whalley:2005nh} is now a separate dynamically
      loadable module \texttt{ThePEGLHAPDF.so} rather than being part 
      of \texttt{libThePEG.so}.
\item The irrelevant three-body decays are no longer included by default when using
      the Randall-Sundrum model.
\item The \ThePEGClass{LeptonLeptonRemnant} class has been renamed
      \ThePEGClass{UnResolvedRemnant} as it is now used in other applications,
      for example in photon radiation from protons in hadronic collisions.
\item The handling of weighted events in the internal \ThePEGClass{AnalysisHandler}
      classes has been improved.
\item Various changes have been made to allow \HWPPClass{Interpolator} objects
      to be written to a persistent \ThePEG\ stream.
\item The obsolete \HWPPClass{BSMCascadeAnalysis} has been removed.
\item A value of the Fermi constant, $G_F$, is now available in the 
      \ThePEGClass{StandardModelBase} class. This allowed the removal
      of several local values of this parameter.
\item The handling of initial-state radiation when the incoming hadron
      is not along the $z$-axis has been improved.
\item The tolerance parameter used to check momentum conservation in the
      \HWPPClass{BasicConsistency} analysis class can now be changed by the user.
\item A new switch has been added to the \HWPPClass{ModelGenerator} class so
      that the branching
      ratios it calculates in BSM models can be outputted in the SLHA format.
\end{itemize}

\pagebreak[3]
The following bugs have been fixed:
\begin{itemize}
\item The example input files for Powheg processes now set the NLO $\alpha_S$ correctly, 
      and are run as part of \texttt{make check}.
\item A problem that led to the truncated shower not being applied in some cases
      has been fixed.
\item An accidental duplication in the calculation of event shapes was removed,
      they are now only calculated once per event. Several other minor issues
      in the event shape calculations have also been fixed.
\item An initialization problem in the internal MRST PDFs was fixed.
\item The scale in the \textsf{Vertex} classes can now be zero where physically possible.
\item The \HWPP\ main program now correctly treats the \texttt{-N} flag as optional.
\item The accuracy of boosts in the $z$-direction has been improved to fix problems with 
      extremely high $p_T$ partons.
\item A bug in the implementation of the PDF weight in initial-state $\bar q\to \bar q g $ splittings has been fixed.
\item A bug in the $\tilde{\chi}^\pm\tilde{\chi}^0W^\mp$ and charged Higgs-sfermions 
      vertices has been fixed.
\item The longitudinal boost of the centre-of-mass frame in hadronic
      collisions is correctly accounted for now in the generation of QED radiation.
\item Numerical problems have been fixed, which appeared in the rare case
      that the three-momenta of the decay products in two-body decays are
      zero in the rest frame of the decay particle.
\item The numerical stability in the \HWPPClass{RunningMass} and
      \HWPPClass{QTildeReconstructor} classes has been improved.
\item The stability of the boosts in the \textsf{SOPTHY} code for the
      simulation of QED radiation has been improved.
\item A problem with forced splittings in the Remnant was fixed.
\item A problem with emissions from antiquarks in the soft matrix element
      correction in \mbox{$e^+e^-\to q \bar q$} was fixed. This also required a
      new tune of the shower and hadronization parameters. 
\item The matrix element correction for QCD radiation in $W^\pm$ decays, which was not
      being applied, is now correctly used in $W^\pm$ decays.
\item The presence of top quark decay modes in SLHA files is now handled correctly.
\item Additional protection against problems due to the shower reconstruction leading to 
      partons with $x>1$ has been added.
\item Changes have been made to allow arbitrary ordering of the outgoing particles in
      BSM processes.
\item Two bugs involving tau decays have been fixed. The wrong masses
      were used in the \HWPPClass{KPiCurrent} for the scalar form factors and
      a mistake in the selection of decay products lead to $\tau^-\to\pi^0K^-$
      being generated instead of $\tau^-\to\eta K^-$.
\item To avoid crashes, better protection has been introduced for the
      case where diquarks cannot be formed from the quarks in a
      baryon-number violating process. In addition, the parents of the
      baryon-number violating clusters have been changed to avoid
      problems with the conversion of the events to \textsf{HepMC}.
\item A bug in the \HWPPClass{QEDRadiationHandler} class that resulted in
      no QED radiation being generated in $W^-$ decays has been fixed.
\item A number of minor fixes to the SUSY models have been made.
\item A fix for the direction of the incoming particle in the calculation
      of two-body partial widths in BSM models has been made.
\item The \textsf{LoopTools} cache is now cleared more frequently to reduce
      the amount of memory used by the program.
\item Negative gluino masses are now correctly handled.
\item A problem with mixing matrices that are not square has been fixed. 
\item A problem in the matrix element correction in $e^+e^-\to t\bar{t}$ events
      has been fixed.
\item The \HWPPClass{MEee2gZ2ll} has been fixed to only include the photon
      exchange diagram once rather than twice as previously.
\item A problem has been fixed that occurred if the same particle was included in the
      list of \HWPPParameter{DecayConstructor}{DecayParticles}.
\item A number of minor problems in the vertices for the UED model have been fixed.
\item The missing identical-particle symmetry factor in \HWPPClass{MEPP2GammaGamma}
      has been included.
\item A floating point problem in the matrix element correction for top decays
      has been fixed.
\end{itemize}

\section{Summary}

  \HWPP\,2.5 is the seventh version of the \HWPP\ program with a complete simulation of 
  hadron-hadron physics and contains a number of important improvements
  with respect to the previous
  version. The program has been extensively tested against
  a large number of observables from LEP, Tevatron and B factories.
  All the features needed for realistic studies for 
  hadron-hadron collisions are now present and  we look forward to 
  feedback and input from users, especially
  from the Tevatron and LHC experiments.

  Our next major milestone is the release of version 3.0, which will be at least as
  complete as \HW\ in all aspects of LHC and linear-collider simulation.
  Following the release of \HWPP\,3.0, we expect that support for the 
  {\sf FORTRAN} program will cease.

\section*{Acknowledgements} 

This work was supported by Science and Technology Facilities Council and the
European Union Marie Curie Research Training Network MCnet under
contract MRTN-CT-2006-035606. SG, SP, CAR and AS acknowledge support
from the Helmholtz Alliance ``Physics at the Terascale''. We would like to thank all those who have
reported issues with the previous release, in particular 
B.~Allanach, A.~Buckley, C.~Gwenlan, K.~Rolbiecki, and J.~Tattersall.
  
\bibliography{Herwig++}
\end{document}